\documentclass[twocolumn,prl,amsmath,amssymb,floatfix,superscriptaddress]{revtex4-1}
\usepackage{color}
\usepackage{graphicx}
\usepackage{dcolumn}
\usepackage{bm}
\usepackage{times}
\begin{document}


\title{Evidence for local moment magnetism in superconducting
FeTe$_{0.35}$Se$_{0.65}$}

\author{Zhijun~Xu}
\affiliation{Condensed Matter Physics and Materials Science
Department, Brookhaven National Laboratory, Upton, New York 11973,
USA} \affiliation{Department of Physics, City College of New York,
New York, New York 10033, USA}
\author{Jinsheng~Wen}
\affiliation{Condensed Matter Physics and Materials Science
Department, Brookhaven National Laboratory, Upton, New York 11973,
USA}\affiliation{Department of Materials Science and Engineering,
Stony Brook University, Stony Brook, New York 11794, USA}
\author{Guangyong~Xu}
\affiliation{Condensed Matter Physics and Materials Science
Department, Brookhaven National Laboratory, Upton, New York 11973,
USA}
\author{Songxue~Chi}
\affiliation{NIST Center for Neutron Research, National Institute of
Standards and Technology, Gaithersburg, Maryland 20899, USA}
\author{Wei Ku}
\author{Genda~Gu}
\author{J.~M.~Tranquada}
\affiliation{Condensed Matter Physics and Materials Science
Department, Brookhaven National Laboratory, Upton, New York 11973,
USA}
\date{\today}

\begin{abstract}
The nature of the magnetic correlations in Fe-based superconductors remains a 
matter of controversy.  To address this issue, we use inelastic neutron 
scattering to characterize the strength and temperature dependence of 
low-energy spin fluctuations in FeTe$_{0.35}$Se$_{0.65}$ ($T_c \sim 14$~K).  
Integrating magnetic spectral weight for energies up to 12 meV, we find a 
substantial moment ($\agt 0.26~\mu_B/$Fe) that shows little change with 
temperature, from below $T_c$ to 300~K.   Such behavior cannot be explained 
by the response of conduction electrons alone; states much farther from the 
Fermi energy must have an instantaneous local spin polarization.  
It raises interesting questions regarding the formation of the spin gap and 
resonance peak in the superconducting state.

\end{abstract}

\pacs{74.70.Xa, 75.25.-j, 75.30.Fv, 61.05.fg}

\maketitle

Antiferromagnetism and superconductivity are common to the phase diagrams of 
cuprate and Fe-based superconductors, and it is frequently proposed that 
magnetic correlations are important to the mechanism of electron 
pairing \cite{mazi10,pagl10}.  Experiments on various Fe-based superconductors 
have demonstrated that magnetic excitations coexist with, and are modified by, 
the superconductivity.
In particular, the low energy spin excitation spectrum is modified by the 
emergence of a ``resonance'' peak and spin gap in the superconducting
phase~\cite{Christianson2008,Lumsden2009prl,Chis2009prl,Inosov2010nf,
Qiu2009,Lumsden2010nf,Wen2010H,Lee2010,Lis2010}. 
Despite some variation in the magnetic structure of the
parent compounds, in all known Fe-based superconductors, the
``resonance'' occurs at the same $\textbf{Q}_{0}\sim(0.5,0.5,0)$ (in
2-Fe unit cell unit), including the $A$Fe$_2$As$_2$
(``122'',$A$=Ba,Sr,Ca)
system~\cite{Christianson2008,Zhaoj2009nf,Chen2009epl,Inosov2010nf,Chis2009prl,Lumsden2009prl},
the $R$FeAsO (``1111'', $R$=La,Ce,Pr,Nd,Gd,Sm)
system~\cite{Clarina2008,Zhaoj2008nm,Huang2008prb,Lee2008jpsj,
Zhaoj2008prb,Qiu2008prb}, and the FeTe$_{1-x}$Se$_{x}$ (``11'')
system~\cite{Qiu2009,Lumsden2010nf,Wen2010H,Lee2010,Lis2010}.  At low
temperature, the resonance is also accompanied by a well-defined but
anisotropic dispersion~\cite{Lee2010,Lis2010} along the transverse
direction, with a spin gap below which there is no spectral weight
in the superconducting state, resembling the spin excitations in
many high T$_c$ cuprates~\cite{Hayden2004,Vignolle2007,Hinkov2007np,Xug2009np}.

One essential and currently unsettled
issue 
is the nature of the magnetism in the Fe-based superconductors~\cite{pagl10}.  In
contrast to the Mott-insulating parent compounds of the
cuprates, the parent compounds of all of the Fe-based superconductors
are poor metals.  This naturally leads to the suggestion of
itinerant magnetism resulting from the nesting of the Fermi surface, or
more generally, enhancement of non-interacting susceptibility
~\cite{Mazin2008,*Kuroki2008,*Scalapino2008PRL,*Chubukov2010PRB}.  Disregarding the apparent failure of such itinerant
picture in producing the so-called bi-collinear magnetic structure
of Fe$_{1+y}$Te \cite{Baow2009prl,*Lis2009}, the spin-fluctuation picture of
superconductivity ~\cite{Mazin2008,*Kuroki2008,*Scalapino2008PRL,*Chubukov2010PRB} is qualitatively appealing, and appears to give a natural explanation for the spin resonance and spin gap \cite{Maier2009}. Nevertheless, there are recent theoretical analyses that suggest that there may be a significant local-moment character to the magnetism \cite{Mazin2009,*Kou2009epl,*Medici2010,*Yinw2010,*Arita2}.
Thus, it is timely to test experimentally whether the weak-coupling approach can quantitatively account for magnetic correlations in the Fe-based superconductors.

In this letter, we report an inelastic neutron scattering study on the temperature
evolution of the low-energy magnetic excitation of an FeTe$_{1-x}$Se$_{x}$
sample with x=65\%.  The
magnetic excitations below $T_c\sim 14$~K are almost identical to
those measured  previously on superconducting FeTe$_{1-x}$Se$_{x}$
samples with $x \alt50\%$ \cite{Qiu2009,Lumsden2010nf,Wen2010H,Lee2010,Lis2010}, having a spin gap of $\sim5$ meV and a resonance at $\sim7$ meV, with anisotropic
dispersion along the direction transverse to ${\bf Q}_0$.
On heating to $T=25$~K, the resonance disappears, with spectral weight
moving into the gap, and the dispersion resembles an ``hour-glass''
shape like those observed in the
cuprates~\cite{Hayden2004,Vignolle2007, Hinkov2007np,gxu2007lbco}. With further
heating,  the spin excitations near the saddle point (5~meV) start
to split in {\bf Q} and become clearly incommensurate, exhibiting a
``waterfall'' structure at 100~K and above, similar to the situation
in underdoped YBa$_2$Cu$_3$O$_{6+x}$ \cite{Hinkov2007np}. However, the
integrated spectral weight below $\hbar\omega=12$~meV remains almost
unchanged as a function of temperature, indicating a large energy
scale associated with the stability of the instantaneous magnetic moment. The absolute
normalization of the low-energy weight gives a lower limit (not counting
the strong spectral weight at higher energies~\cite{Lumsden2010nf}) of
the magnetic moment per Fe site to be $\sim 0.26 \mu_B/$Fe. Such a
robust and sizable moment is apparently beyond the standard
consideration of spin-density-wave picture \cite{Mazin2008,*Kuroki2008,*Scalapino2008PRL,*Chubukov2010PRB}, and strongly suggests
that local moment magnetism is present (and likely dominant) in the
Fe-based superconductors \cite{Mazin2009,*Kou2009epl,*Medici2010,*Yinw2010}.

\begin{figure}[ht]
\includegraphics[width=0.9\linewidth]{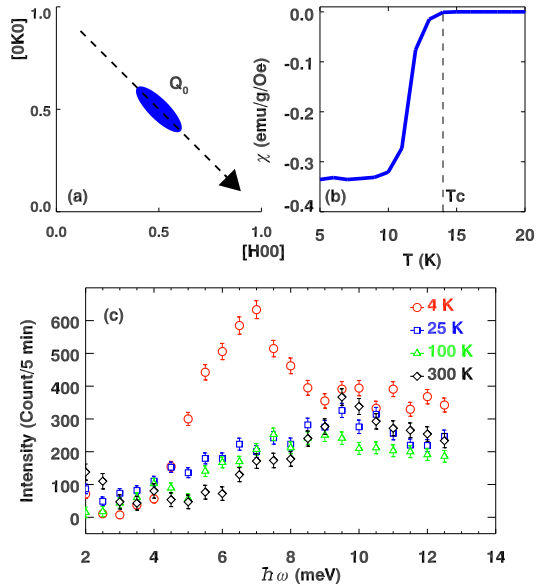}
\caption{(Color online) (a) The schematic diagram of the neutron
scattering measurements in the $(HK0)$ zone. Dashed lines denote
linear scans performed across $\textbf{Q}_{0}=(0.5,0.5,0)$ in the
text. (b) ZFC magnetization measurements by SQUID with a
5~Oe field perpendicular to the $a$-$b$ plane. $T_{c}\sim$14K is
marked by a dash line. (c) Constant Q scans at $\textbf{Q}_{0}$
taken at different temperatures: 5~K (red circles), 25~K (blue
squares), 100~K (green triangles), and 300~K (black diamonds). Fitted
background obtained from constant energy scans has been subtracted
from all data sets.} \label{fig:1}
\end{figure}

The single-crystal sample used in the experiment was grown by a
unidirectional solidification method with nominal composition of
Fe$_{0.98}$Te$_{0.35}$Se$_{0.65}$ (8.6g). The bulk susceptibility, measured with a superconducting quantum interference device (SQUID) magnetometer, is shown in 
in Fig.~\ref{fig:1} (b),  indicating $T_{c}\sim14~K$. Neutron scattering experiments were carried
out on the triple-axis spectrometer BT-7 located at the NIST Center
for Neutron Research. We used beam collimations of open-$50'$-S-$50'$-open 
(S = sample) with fixed final energy of 14.7~meV
and two pyrolytic graphite filters after the sample. The lattice
constants for the sample are $a = b = 3.81$~\AA, and $c = 6.02$~\AA,
using a unit cell containing two Fe atoms.
The inelastic scattering measurements have been performed in the
$(HK0)$ scattering plane [Fig.~\ref{fig:1} (a)]. The data are
described in reciprocal lattice units (r.l.u.) of $(a^*, b^*, c^*) =
(2\pi/a, 2\pi/b, 2\pi/c)$. Absolute normalizations are performed
based on measurements of incoherent elastic scattering from the sample.

\begin{figure}[ht]
\includegraphics[width=\linewidth]{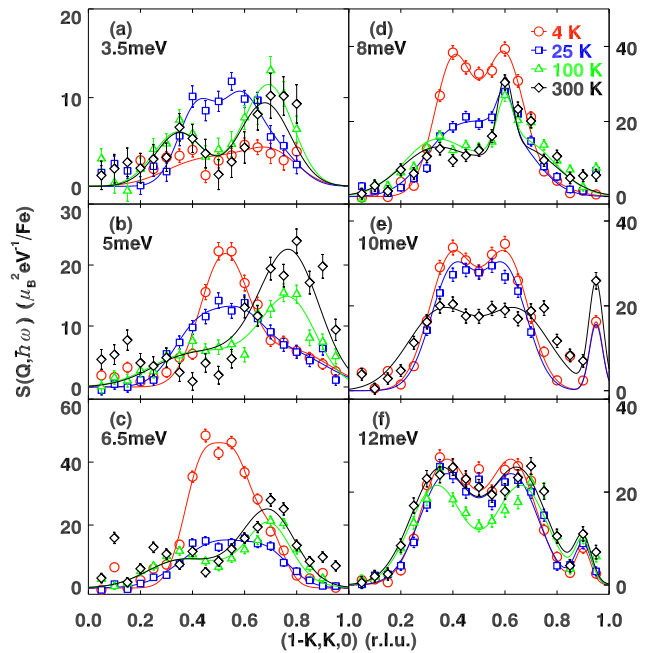}
\caption{(Color online) Constant energy scans at $(1-K,K,0)$ with
different temperatures: 4~K (red circles), 25~K (blue squares), 100~K
(green triangles) and 300~K (black diamonds) at different
$\hbar\omega$: (a) 3.5~meV, (b) 5~meV, (c) 6.5~meV, (d) 8~meV, (e)
10~meV, and (f) 12~meV. A flat fitted background has been subtracted
from all data sets. The solid lines are based on the fit described
in the text. The error bars represent the square root of the number
of counts.} \label{fig:3}
\end{figure}

\begin{figure}[ht]
\includegraphics[width=\linewidth]{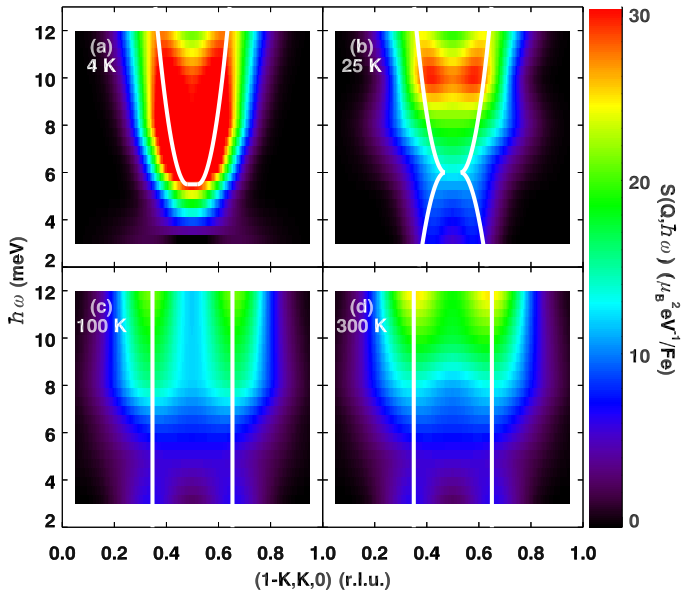}
\caption{(Color online) Contour intensity maps showing the fitted magnetic
scattering intensity versus $\hbar\omega$ and {\bf Q} at different temperatures:
(a) 4~K, (b) 25~K, (c) 100~K and (d) 300~K. } \label{fig:4}
\end{figure}

\begin{figure}[ht]
\includegraphics[width=0.9\linewidth]{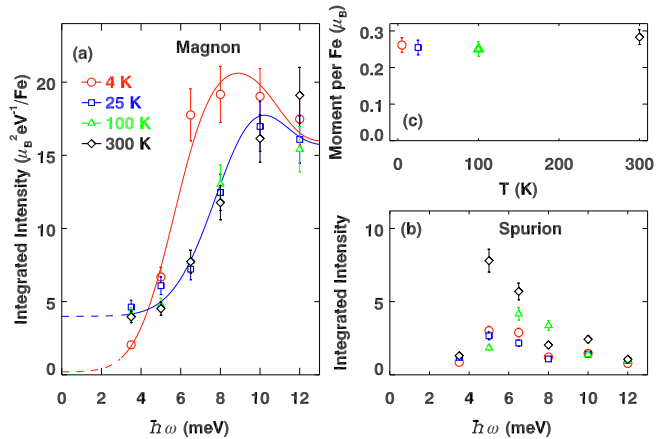}
\caption{(Color online) (a) {\bf Q}-integrated (integrated only in
one-dimension, along the transverse direction) magnetic intensity,
obtained based on the fit described in the text, plotted vs.
temperature. (b) {\bf Q}-integrated intensity for the spurious peak around
(0.25,0.75,0), plotted vs. temperature. (c) Magnetic moment per Fe
site vs.\ temperature.} \label{fig:5}
\end{figure}

Low energy spin excitations are mainly distributed near
$\textbf{Q}_{0}$ in-plane wave-vector, similar to the case in the
50\% Se doped 
sample~\footnote{There is no elastic magnetic intensity found near $(0.5,0,0.5)$, which is the antiferromagnetic ordering wave vector in the parent compound, and very little spectral weight near $(0.5,0,0)$ at low energies and low temperature.}\cite{zxu2010fetese1}. In
Fig.~\ref{fig:1}(c), we show constant-{\bf Q} scans at $\textbf{Q}_{0}$
from 4~K to 300~K. There is a clear resonance peak for data
taken in the superconducting phase ($T=4$~K, red circles). When
heated above $T_c$, the resonance peak disappears, and spectral
weight starts to fill in the gap below $\Delta \sim 5$~meV. For the
normal state, the intensity at $\textbf{Q}_{0}$ appears to peak at
around $\hbar\omega \sim 10$~meV. These results are in good
agreement with previous neutron scattering
measurements~\cite{Lee2010,Qiu2009}, indicating that further Se
doping above the optimal value of 50\%\ does not significantly alter
the low-energy magnetic excitations in the system.

Constant energy scans across $\textbf{Q}_{0}$, performed in the
transverse direction, are plotted in Fig.~\ref{fig:3}.  One can see
how the resonance disappears with heating in Fig.~\ref{fig:3} (c) and (d). For $\hbar\omega \le 6.5$~meV, Fig.~\ref{fig:3} (a)-(c), we note that
the peak on the right side [larger $K$ side, near $(0.25,0.75,0)$] is
further out in {\bf Q}, with respect to $Q_0$, compared to its
counter-part on the left (small $K$) side, and becomes
disproportionately strong. This behavior is inconsistent
with crystal symmetry, which magnetic or simple phonon
scattering must follow. The nature of this spurious peak is
not entirely known. It is very likely not associated with magnetic
scattering from the sample; its growth with temperature suggests that it arises  from multi-scattering
processes involving certain phonon modes. Fortunately, it only appears on the large $K$ side, leaving the small $K$ side uncontaminated. In
our data analysis, we fit the magnetic signal using a double Gaussian
function, with two peaks split symmetrically about ${\bf Q}_0$, plus a single Gaussian function for the spurious peak.
The fitted magnetic intensities are presented as contour maps in
Fig.~\ref{fig:4}. With the spurious peak removed, one can easily see the
evolution of the magnetic excitation spectrum with temperature.

In the superconducting phase, Fig.~\ref{fig:4}(a), there is
very little spectral weight below 5~meV, while the excitations disperse outwards at higher energies.    As a function of temperature [Fig.~\ref{fig:4}(a)-(d)], the dispersion at the highest energies changes little, and one can
still observe well defined magnetic excitations at $\hbar\omega =
12$~meV up to $T=300$~K.
The temperature effect on the dispersion below the resonance energy is
much more pronounced.
On warming from 4~K to 25~K,  intensity that emerges below the gap appears to disperse outwards slightly, as shown
in Fig.~\ref{fig:4} (b). Our results
are consistent with those in in Ref.~\onlinecite{Lis2010},
where the spectrum is narrowest in {\bf Q} at the saddle point around 5~meV, and
becomes broader for energy transfers both above and below for $T>T_c$.

With further heating, the Q-dependence of the spectrum changes most
dramatically near the saddle point.
At $T=100$~K, the lower part of the dispersion clearly moves outwards from
$\textbf{Q}_{0}$, as shown in Fig.~\ref{fig:4}(c).
The saddle point at 5~meV actually disappears, and the dispersion
becomes clearly incommensurate and almost vertical.
There is little change between 100~K and 300~K.
The change in dispersion from 4~K to 100~K is qualitatively similar to behavior reported for underdoped YBa$_2$Cu$_3$O$_{6+x}$ \cite{Hinkov2007np}.

In Fig.~\ref{fig:5} (a) and (b), we plot the intensities, integrated along ${\bf Q}=(1-K,K,0)$,
of the magnetic scattering and the spurious peak. The effect of the resonance in
the superconducting phase is observable up to $\hbar\omega \sim
10$~meV. The plot
of the spurious-peak  intensity shows signs of temperature
activation, and is peaked near 5 meV; in any case, its scale is generally small compared to the magnetic signal.

Our key result is obtained by integrating the magnetic signal over {\bf Q} and $\hbar\omega$.  For the {\bf Q} integration, we
assume the peak width along the longitudinal direction is same as
transverse direction and that the response is uniform along $L$.  For energy, we integrated over the interval 0~meV$\le\hbar\omega\le12$~meV, using the low-energy extrapolation indicated by the dashed lines in Fig.~\ref{fig:5}(a).  From this integral, we obtain an instantaneous magnetic moment of 0.26(7)~$\mu_B$ per Fe.  The temperature dependence of this quantity is negligible, as shown in Fig.~\ref{fig:5}(c).

The moment we have evaluated is only a fraction of the total moment per site, considering that
previous measurements have shown significant spectral weight all the
way up to a few hundred meV~\cite{Lumsden2010nf}. Nevertheless, such
a large low-energy magnetic response is already an order of
magnitude larger than what is expected from a simple itinerant picture.  
For example, the density of states at the Fermi energy ($E_{\rm F}$) has been 
calculated to be $\sim1.5$~ev$^{-1}$ per Fe for FeSe \cite{Lee2008}.  If we 
assume that the electronic states within an energy range of 12 meV are fully 
spin polarized, then obtain an estimated moment of $\sim0.02$~$\mu_B$.  Of 
course, photoemission studies have reported band renormalizations in the range 
of 3 to 20 \cite{Tamai2010,Zhang2010}, which would substantially increase the 
available density of states at $E_{\rm F}$; however, one must also consider the 
cause of the renormalizations.  In theory, one can include interactions that 
enhance the magnetic response using the random phase approximation; however, 
at least one attempt to do this~\cite{Kariyado2009}  has found that the strength 
of the low-energy 
magnetic weight is strongly temperature dependent, in contrast to our 
experimental result.  The observed lack of temperature dependence suggests that 
electronic states over a large energy range contribute to the effective moment, 
which is consistent with having a significant local moment, as suggested by 
recent theoretical work \cite{Mazin2009,*Kou2009epl,*Medici2010,
*Yinw2010,*Arita2}.

This leads to an interesting question.  For the itinerant picture, the spin gap and resonance come out naturally from the pairing gap for the quasiparticles---although they are sensitive to the symmetry of the order parameter.  If the magnetic moments involve states at high binding energies, then one must reconsider the evaluation of the resonance.  It is clear that the magnetic correlations are sensitive to the development of pairing and superconductivity; however, the electrons involved in the pairing and in the magnetism are not necessarily identical.  Similar issues have been raised in the case of cuprates \cite{Xug2009np}.  These issues also raise questions concerning the nature of the pairing mechanism.

We thank Weiguo Yin and Igor Zaliznyak for useful discussions. This work is
supported  of the Office of Basic Energy Sciences, U.S. Department of Energy under contract No.
DE-AC02-98CH10886.  JSW and ZJX are supported by the same source through the Center for
Emergent Superconductivity, an Energy Frontier Research Center.

%

\end{document}